\documentclass[preprint,tightenlines,superscriptaddress]{revtex4}
\usepackage[utf8x]{inputenc}
\usepackage{amsmath} 
\usepackage{amsfonts,hyperref}
\usepackage{slashed}
\usepackage[dvips]{graphicx}
\usepackage{color}
\usepackage{units}
\usepackage{bbm}
\allowdisplaybreaks

\newcommand{\e}{{\text{e}}}
\renewcommand{\i}{{\text{i}}}
\renewcommand{\e}{{\text{e}}}

\usepackage{color}

\begin{document}
 \author{{\bf Clément Stahl}\email{clement.stahl@icranet.org}}
 \affiliation{ICRANet, Piazzale della Repubblica 10, 65122 Pescara, Italy}
 \affiliation{Dipartimento di Fisica, Universit\`a di Roma “La Sapienza”, Piazzale Aldo Moro 5, 00185 Rome, Italy}
 \affiliation{Universit\'e de Nice Sophia Antipolis, 28 Avenue de Valrose, 06103 Nice Cedex 2, France}
   \affiliation{Nordita, KTH Royal Institute of Technology and Stockholm University,Roslagstullsbacken 23 SE-106 91 Stockholm, Sweden}
 \author{{\bf She-Sheng Xue}\email{xue@icra.it}}
 \affiliation{ICRANet, Piazzale della Repubblica 10, 65122 Pescara, Italy}
 \affiliation{Dipartimento di Fisica, Universit\`a di Roma “La Sapienza”, Piazzale Aldo Moro 5, 00185 Rome, Italy}

\date{\today}

\title{Schwinger effect and backreaction in de Sitter spacetime.}

\begin{abstract}
 We consider the particle-antiparticle pairs produced by both a strong electric field and de Sitter curvature. We investigate in 1 + 1 D the backreaction of the pairs on the electromagnetic field. To do so we describe the canonical quantization of an electromagnetic field in de Sitter space and add in the Einstein-Maxwell equation the fermionic current induced by the pairs. After solving this equation, we find that the electric field gets either damped or unaffected depending on the value of the pair mass and the gauge coupling. No enhancement of the electromagnetic field to support a magnetogenesis scenario is found. The physical picture is that the Schwinger pairs locally created screen the production and amplification of the electromagnetic field. However, if one considers light bosons created by the Schwinger mechanism, we report a solution to the Einstein-Maxwell equation with an enhancement of the electromagnetic field. This solution could be a new path to primordial magnetogenesis.
\end{abstract}

\maketitle

%%%%%%%%%%%%%%%%%%%%%%%%%%%%%%%%%%%%%%%%%%%%%%%%%%%%%%%%%%%%%%%%%%%%%%%%%%%%%%%%%%%%%%%%%%%%%%%%%%%%%%%%%%%%%%%%%%%%%%%%%%%%%%%%%%%%
%%%%%%%%%%%%%%%%%%%%%%%%%%%%%%%%%%%%%%%%%%%%%%%%%%%%%%%%%%%%%%%%%%%%%%%%%%%%%%%%%%%%%%%%%%%%%%%%%%%%%%%%%%%%%%%%%%%%%%%%%%%%%%%%%%%%
\par Particle production from the vacuum is an important topic in modern theoretical physics. On the one hand, the Schwinger effect \cite{Sauter1931,Heisenberg1936,Schwinger1954A} (\emph{ie.}~excitation of particle-antiparticle pairs tunneling out of the Dirac sea \cite{Ruffini:2009hg,Gelis:2015kya}) is one thrilling playing field for theoretician because of its non-perturbative nature in the electric field. On the other hand, the gravitational pair creation (during inflation \cite{Baumann:2014nda}, the Hawking radiation near black holes \cite{Hawking:1974sw} and the Unruh radiation for accelerating observers \cite{Unruh:1976db}) is a well known phenomenon. However despite 50 years of theoretical existence, no pairs have been directly detected today. The main hope of detection, for Schwinger effect will come with the next generation of high energy laser\footnote{European X-Ray Free-Electron Laser [XFEL], Extreme Light Infrastructure [ELIA], High Power Laser Energy Research [HiPer]} which will be operational within 5 years. Regarding the gravitational pair creation, the main hope is to detect the celebrated B-modes of the cosmic microwave background anisotropies \cite{Bucher:2015sbn}. Awaiting for these possible discoveries, a rising field of research is dedicated to include in one unique system both the electric and gravitational pair creation. Having both a strong electric and gravitational field could be realized in early universe physics in a phase of accelerating expansion. During this phase, both electromagnetic and a test particle fields are enhanced due to a parametric amplification of vacuum fluctuations.
\par In the past years, the Schwinger effect in de Sitter spacetime ($\text{dS}_n$) was investigated for both fermions and bosons in various space dimensions \cite{Frob:2014zka,Kobayashi:2014zza,Fischler:2014ama,Stahl:2015gaa,Bavarsad:2016cxh,Hayashinaka:2016qqn,Hayashinaka:2016dnt}. In all those studies, the backreaction of the pairs has been always neglected. However, such backreaction could be important. For instance, already in flat spacetime, for 1 + 1 D fermions, results have been shown and the phenomenon of plasma oscillation has been discovered  \cite{Kluger:1991ib,PhysRevD.45.4659} (see also \cite{Ruffini:2003cr,Ruffini:2007jm}). We aim at generalizing these results to the Schwinger effect in $\text{dS}_n$. The point of view taken in this letter is to ignore the transverse direction of the electric field to make the system simpler and grasp some physical insight on the process. Furthermore, from the study of the Schwinger effect in $\text{dS}_2,\text{dS}_3$ and $\text{dS}_4$, the behavior of the created particles with respect to the electric and gravitational strength is similar \cite{Bavarsad:2016cxh}.
\paragraph*{QED in $\text{dS}_2$} We consider the Lagrangian for 1 + 1 D QED minimally coupled to gravity:
\begin{align}
\label{eq:action}
S=\int\text{d}^2 \mathcal{L}=\int \text{d}^2 x  \sqrt{-g(x)} \left[-\frac{1}{4} g^{\alpha \mu}(x) g^{\beta \nu}(x) F_{\mu \nu}(x)F_{\alpha \beta}(x) -j^{\mu}(x)A_{\mu}(x) \right],
\end{align}
where $j^{\mu}(x)$ is the fermionic current and the field strength is defined with the potential vector: $F_{\mu \nu}(x)\equiv \partial_{\mu}A_{\nu}(x)-\partial_{\nu}A_{\mu}(x)$. The electromagnetic field is assumed to be a “test” field on the background $\text{dS}_2$:
\begin{equation}
ds^2 = a^2(\eta)(d\eta^2-dx^2), \hspace{1cm} -\infty < \eta < 0
\end{equation}
where the scale factor is $a(\eta)=-\frac{1}{H\eta}$. In unit of mass, in 1 + 1 D, the gauge field $A_{\mu}$ is dimensionless and the U(1) gauge coupling constant $e$ has dimension 1. We assume the temporal gauge: $A_0(x)=0$ and will drop from now on the subscript and write $A(x)$ for $A_1(x)$. 
In 1 + 1 D there is no magnetic field and the electric field is a pseudoscalar and the momentum conjugate to $A(x)$, in $\text{dS}_2$ it reads: 
\begin{equation}
E(\eta,x)=-\frac{1}{a}A'(\eta,x),
\label{eq:electric}
\end{equation}
where a prime denotes a derivative with respect to the conformal time.
%%%%%%%%%%%%%%%%%%%%%%%%%%%%%%%%%%%%%%%%%%%%%%% La quantisation %%%%%%%%%%%%%%%%%%%%%%%%%%%%%%%%%%%
%%%%%%%%%%%%%%%%%%%%%%%%%%%%%%%%%%%%%%%%%%%%%%%%%%%%%%%%%%%%%%%%%%%%%%%%%%%%%%%%%%%%%%%%%%%%%%%%%%%%%%%%%%%%%%%%%%%%%%%%%%%%%%%%%%%%%%%%%%%%%%%%%%%%%%%%%%%%%%%%%%%%%%%%%%%%%%%%%%%%%%%%%%%%%%%%%%%%%%
%%%%%%%%%%%%%%%%%%%%%%%%%%%%%%%%%%%%%%%%%%%%%%% La quantisation %%%%%%%%%%%%%%%%%%%%%%%%%%%%%%%%%%%
The canonical commutation relation between $A(x)$ and its canonical momentum $\pi(x)=\frac{1}{a(\eta)}E(\eta,x)$ is then imposed:
\begin{equation}
\left[A(\eta,x),\pi(\eta,y)\right]=\i \delta(x-y),
\label{eq:comrelation}
\end{equation}
the others commutators being zero, $A(\eta,x)$ is then promoted to be the potential vector operator and can be Fourier expanded:
\begin{equation}
\label{eq:Fourier}
\hat{A}(\eta,x)=\int \frac{d k}{(2\pi)^{\frac{1}{2}}}\left(A(\eta,k)\hat{b}(k)e^{\i k.x}+\hat{b}^{\dagger}(k)A^*(\eta,k)e^{-\i k.x}\right).
\end{equation}
The creation and annihilation operators are $\hat{b}_p(k)$ and $\hat{b}^{\dagger}_p(k)$ with $k$ the comoving wavenumber. In order to satisfy Eq.~(\ref{eq:comrelation}), the creation and annihilation operators must satisfy the following commutation relations:
\begin{equation}
\left[\hat{b}(k), \hat{b}^{\dagger}(k')\right]=\delta(k-k'),
\end{equation}
the others commutators being zero. The Wronskian condition reads:
\begin{equation}
A(\eta,k)A'^*(\eta,k)-A'(\eta,k)A^*(\eta,k)=\i.
\end{equation}
%%%%%%%%%%%%%%%%%%%%%%%%%%%%%%%%%%%%%%%%%%%%%%%%%%%%%%%%%%%%%%%%%%%%%%%%%%%%%%%%%%%%%%%%%%%%%%%%%%%
Using \ref{eq:electric} together with \ref{eq:Fourier}, one finds the Fourier transform of the electric field:
\begin{equation}
E(\eta,k)=-\frac{1}{a}A'(\eta,k).
\label{eq:electricF}
\end{equation}
Adopting the expression for the fermionic current for the case of a constant electric field given in equation (96) of \cite{Stahl:2015gaa} together with \ref{eq:electricF}, we obtain the following expression for the current:
\begin{align}
& j^{1}(\eta,k)=
\frac{e}{\pi}\omega_F(\eta,k)\frac{\sinh\left(\frac{2 \pi}{H}e\eta A'(\eta,k)\right)}{\sinh\left(\frac{2\pi}{H}\omega_F(\eta,k)\right)}, \label{eq:current} \\
&\omega_F(\eta,k) \equiv \sqrt{e^2\eta^2 A'(\eta,k)^2+m^2}. \label{eq:ferm}
\end{align}
By taking this expression, one does a local constant electric field approximation and assumes the expression for the Schwinger effect derived for constant electric field is valid at every time instance of the evolution of the electric field. This approximation is valid for two reasons:
\begin{itemize}
\item The population of pairs is always dominated by the one created within a Hubble time.
\item The typical time scale of the pair creation by Schwinger effect ($t_{\text{pairs}} \sim \frac{1}{H}$) is much larger than the typical time scale of the evolution of the electric field ($t_E \sim \frac{1}{k}$). 
\end{itemize}
Those considerations transcribe into the condition $\frac{k}{H} \gg 1$, for which our results are valid. We furthermore emphasize that in this regime, the electromagnetic field is not sensitive any curvature effect because its wavelength is greater than the curvature radius implying that deep in the ultraviolet regime ($\frac{k}{H} \gg 1$) the electric field propagates as in Minkowski spacetime. In the infrared regime ($\frac{k}{H} \ll 1$) the results presented cannot be totally trusted as the Schwinger effect has to be computed to take into account the rapid spatial variations of the electric field. We argue however that the results presented afterward, if not quantitatively exact still have an interesting qualitative behavior. Furthermore, it is known that even in flat spacetime analytic solutions (for Schwinger pair creation) exist only for a few field configurations (constant electric field, Sauter pulse), so the use of the constant electric field as a toy model is a need to obtain insight into this complex problem.

\par The computation of the Schwinger effect in \cite{Stahl:2015gaa} included only the spatial part of the current, however allowing for backreaction, the conservation of the current induces a non-zero $j^0(k)$, which itself induces a second backreaction via the zeroth component of the Maxwell equation. Following \cite{Kobayashi:2014zza}, this approach is not totally consistent with the choice of gauge, as $A_{0}$ cannot be exactly zero in the presence of charge. For a first simple calculation in this letter, however, we consider only the spatial part of the Einstein-Maxwell equation which reads then:
\begin{equation}
A''(\eta,k)+k^2 A(\eta,k)=a^2(\eta) j^1(\eta, k). \label{eq:motion}
\end{equation}
For the initial state, we impose a Bunch-Davies vacuum:
\begin{equation}
\lim_{\eta \to  -\infty} A(\eta,k) =\frac{\e^{-\i k \eta}}{\sqrt{|2k|}}, \hspace{1cm} \lim_{\eta \to  -\infty} A'(\eta,k) = \frac{-\i k}{\sqrt{|2k|}}\e^{-\i k \eta}.
\label{eq:boundary}
\end{equation}
As usually in cosmology, we define the electric power spectrum as:
\begin{align}
& \mathcal{P}_E =\frac{k}{a^2} |A'(\eta,k)|^2.
\label{eq:power}
\end{align}
Observe that the power spectrum may or may not go to zero in the asymptotic future ($\eta \rightarrow 0$) depending on the behavior of the electric field. We hence will also define the ratio of the power spectrum when the backreaction are turned on to the free response (when $j^{\mu}(x)=0$):
\begin{align}
& \frac{\mathcal{P}_E}{\mathcal{P}_E^{\text{free}}}=\frac{|A'(\eta)|^2}{|A'(\eta)|^2_{\text{free}}}, 
\label{eq:ratio}
\end{align}
with $|A'(\eta)|^2_{\text{free}}=|k|/2$.
\paragraph*{Backreaction and plasma oscillation}
Equation (\ref{eq:motion}) does not have an analytical solution  but it is possible to perform a numerical resolution. To do so, we will first reformulate it with the help of dimensionless variables:
\begin{equation}
\epsilon \equiv \frac{e}{H}, \hspace{1cm} \mu \equiv \frac{m}{H}, \hspace{1cm} \tau \equiv \eta H.
\label{eq:dimless}
\end{equation}
Equation (\ref{eq:motion}) becomes then:
\begin{equation}
\ddot{A}(\tau,k)+\frac{k^2}{H^2}A(\tau,k)=\frac{\epsilon}{\pi \tau^2}\omega_F(\tau,k) \frac{\sinh(2\pi \epsilon \dot{A}(\tau,k) \tau)}{\sinh(2\pi\omega_F(\tau,k))},
\end{equation}
where a dot denotes a derivative with respect to the dimensionless time $\tau$ and $\omega_F(\tau,k) \equiv \sqrt{\epsilon^2\dot{A}(\tau,k)^2\tau^2+\mu^2}  $.
\par
\begin{figure}[!htb]
    \centering
     \includegraphics[width=0.5\textwidth]{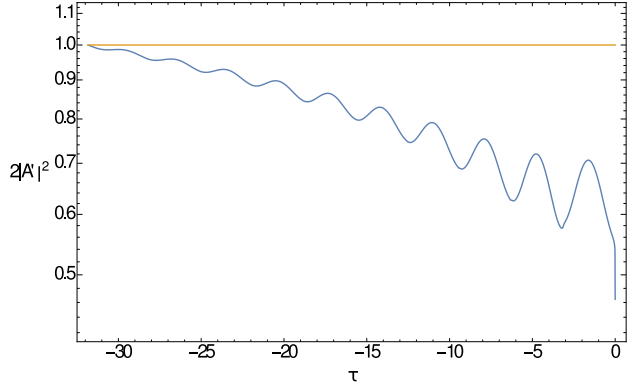} 
 \caption{We propose an example of the behavior of the mode function where $H=k=\epsilon=\mu=1$. The upper yellow constant line is the free case ($j=0$). The blue line is the mode function when the backreaction corrections are turn on. In the asymptotic past, the current vanishes and the two functions are equal. As one approaches the asymptotic future, the plasma oscillations appear.}
 \label{fig:mode function}
\end{figure}

\begin{figure}[!htb]
    \centering
    \includegraphics[width=.5\textwidth]{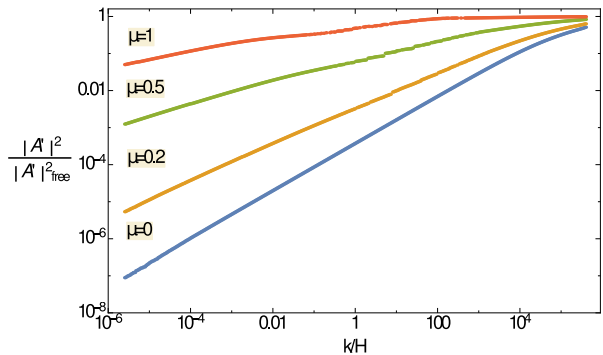} 
 \caption{We plot the power spectrum as given by equation (\ref{eq:ratio}). The selected parameters are: $H=\epsilon=1$. From bottom to top, the values of the mass are: $\mu=0,0.2,0.5,1$}
 \label{fig:varym}
\end{figure}

\begin{figure}[!htb]
        \centering
      \includegraphics[width=.5\textwidth]{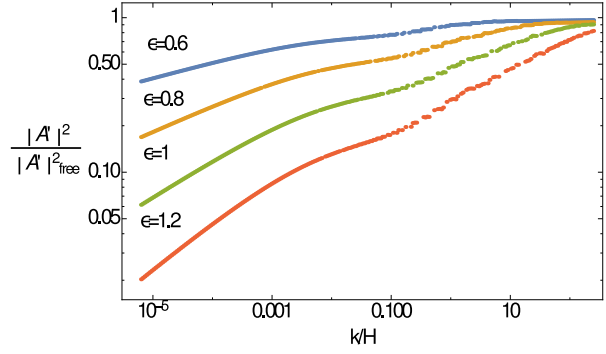} 
 \caption{We plot the power spectrum as given by equation (\ref{eq:ratio}). The selected parameters are: $H=\mu=1$. From top to bottom, the values of the gauge coupling are: $\epsilon=0.6,0.8,1,1.2$. For intermediate values of $k$, numerical instabilities do not allow us to compute the power spectrum for all $k$.}
 \label{fig:varye}
\end{figure}

 We numerically solved equation (\ref{eq:motion}), with the boundary conditions (\ref{eq:boundary}). The result is displayed in figure \ref{fig:mode function} which shows a typical behavior for the mode function $|A'(\tau,k)|^2$. In the asymptotic past, the backreaction term can be neglected. Then one observes plasma oscillations which resemble to the one found in flat spacetime \cite{PhysRevD.45.4659}. However these oscillations are damped by the dilution due to the expansion of the Universe. Qualitatively, the frequency of these plasma oscillations is inversely proportional to the comoving momentum $k$. Indeed, as we will also discuss later, in the ultraviolet regime: for $\frac{k}{H} \gg 1$, the electromagnetic field is not sensitive to gravity and is unaffected by the Schwinger effect. Conversely, in the infrared regime, for $\frac{k}{H} \ll 1$, the plasma oscillations are the dominant physical phenomenon and their frequency is increased.
 Figures \ref{fig:varym} and \ref{fig:varye} show typical power spectra. In the ultraviolet regime ($\frac{k}{H} \gg 1$), the electric power spectrum is unaffected by the Schwinger effect because the backreaction of the current is negligible. Indeed, in this regime, the electromagnetic oscillations are outside the de Sitter horizon and are unaffected by gravity effects. However in the infrared regime ($\frac{k}{H} \ll 1$), the power spectrum gets significantly reduced due to the significant screening of the backreaction of the Schwinger pairs. Observe that varying the Hubble rate $H$ just changes the window of the comoving momentum $k$ considered. Indeed the relevant quantity to understand the backreaction effect is $\frac{k}{H}$.
The current is always positive in the direct space, as a consequence the power spectrum is always reduced with respect to its flat spacetime value. As a result, the electromagnetic field is not significantly enhanced. We will now turn to a phenomenological analysis of the parameters $\epsilon$ and $\mu$ to quantify how much the electric power spectrum is damped.
  \par We are now in the position to phenomenologically discuss the impact of the mass parameter (cf Fig.~\ref{fig:varym}). The more $\mu$ dwindles, the more the power spectrum dwindles. Indeed the heavier the mass of the pairs is, the smaller the current is (cf.~\cite{Stahl:2015gaa}). This implies that less pairs are created so that the screening of the electric field is less important.
\par
Now we turn to the phenomenological study of the gauge coupling $\epsilon$. The more $\epsilon$ increases, the more the power spectrum dwindles. The more the gauge coupling increases, the stronger the photon and pairs are coupled together and the less important the amplification of the electric field is.
\par
\begin{figure}[!htb]
    \centering
     \includegraphics[width=0.5\textwidth]{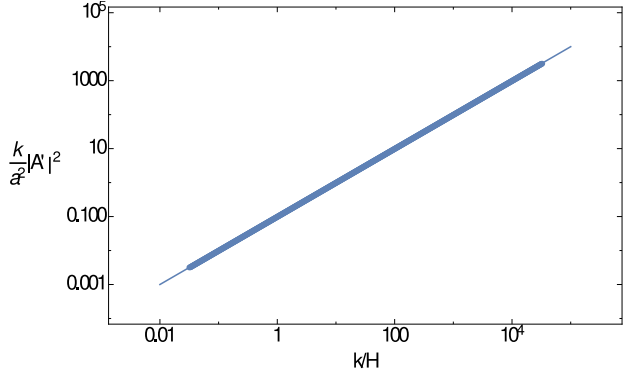} 
 \caption{We report a non-trivial solution where the power spectrum gets enhanced, in this case $\mu=0$ and $\epsilon=1$. The dots are the result of the numerical integration of (\ref{eq:motion}) whereas the plain line is the analytical estimate $|c_{\text{sol}}|^2 k$.} 
 \label{fig:IRHCresul}
    \end{figure}

 In all the scenarii depicted so far, no enhancement of the electromagnetic field is reported and in agreement with \cite{Kobayashi:2014zza}. Namely via backreaction, triggering the Schwinger effect during inflation has the only effect of constraining magnetogenesis scenario.
\paragraph*{Possible amplification of the electromagnetic field.} We now consider bosonic pairs created by the Schwinger effect, in this case the current reads \cite{Frob:2014zka}:
\begin{align}
& j^{1}(\eta, k)=
\frac{e}{\pi}\omega_B(\eta,k)\frac{\sinh\left(\frac{2 \pi}{H}e\eta A'(\eta,k)\right)}{\sinh\left(\frac{2\pi}{H}\omega_B(\eta,k)\right)}, \label{eq:currentb} \\
& \omega_B(\eta,k) \equiv \sqrt{e^2\eta^2 A'(\eta,k)^2+m^2-H^2/4}. \label{eq:pulsb}
\end{align} 
Comparing with equation (\ref{eq:ferm}) the only change is $\omega_B(\eta,k)^2 \rightarrow \omega_F(\eta,k)^2 -H^2/4$. Therefore the parameter space to be examined is $\frac{m^2}{H^2} \in (0,1/4)$. We propose an ansatz $A'(\eta)=\frac{c}{\eta}$ which is the minimal requirement for the solution in order to have enhancement of the electromagnetic field, where the parameter $c$ is to be determined. For infrared modes ($k/H \ll 1$), the current (\ref{eq:currentb}) becomes the dominant term in the Einstein-Maxwell equation (\ref{eq:motion}). Using this ansatz together with (\ref{eq:motion}), (\ref{eq:dimless}), (\ref{eq:currentb}) and (\ref{eq:pulsb}), we find the following scalar equation which may or may not have solution depending on the parameters:

\begin{equation}
-c=\frac{\epsilon}{\pi} \frac{\sinh(2\pi \epsilon c) \omega_c}{\sinh(2\pi \omega_c)},
\label{eq:forc}
\end{equation}
where $ \omega_c \equiv \sqrt{\mu^2+\epsilon^2 c^2-1/4}$. Observe that in the fermionic case, the factor “$-1/4$” is absent, the only trivial solution is $c=0$, no enhancement solution is found. In the bosonic case, the factor “$-1/4$” is present and we consider the case of $\mu =0$. In this case a non-trivial solution of (\ref{eq:forc}) can be found numerically if $c$ is purely imaginary. In Fig.~\ref{fig:IRHCresul}, we plot a typical solution which arises for $\epsilon=1$, in this case $c_{\text{sol}} =  0.32 \i $. We have numerically checked that it is the imaginary part of the electric field $A'(\eta)$ (cf.~Eq.~(\ref{eq:electric})) which is enhanced while the real part stays roughly constant. In the infrared regime ($k/H \ll 1$), using expansion in powers of $k/H$, we get analytical estimates for the power spectrum: $P_E(k) = |c_{\text{sol}}|^2 k$. Both the numerical results and the analytic estimates are plotted in Fig.~\ref{fig:IRHCresul}.
\par To exist, this solution needs to be stable. We look at stability by considering a small perturbation around the solution $A'(\eta)=\frac{c}{\eta}$. We introduce $\alpha(\eta)$ such that $A'(\eta)=\frac{c}{\eta}+\alpha(\eta)$, with $\alpha(\eta) \ll \frac{c}{\eta}$.
Using this ansatz together with (\ref{eq:motion}),(\ref{eq:dimless}), (\ref{eq:currentb}) and (\ref{eq:pulsb}), to the leading order, we find that $\alpha(\eta)$ satisfies the following equation:
\begin{equation}
\alpha'(\eta)=\frac{\alpha(\eta)}{\eta} w,
\end{equation}
whose solution is $\alpha(\eta) \sim \eta^w$ and $w$ given by:
\begin{equation}
w= 2\epsilon^2  \omega_c \frac{\cosh(2\pi c \epsilon)}{\sinh(2 \pi  \omega_c)}+c \epsilon^3\frac{\sinh(2\pi c \epsilon)}{ \pi \omega_c \sinh(2\pi \omega_c)}-2 c \epsilon^3 \frac{\coth(2\pi \omega_c)\sinh(2\pi c \epsilon)}{\sinh(2 \pi \omega_c)}.
\end{equation} 
The solution is stable if $w>0$. In our specific case ($\mu=0$ and $\epsilon=1$), we find $w(c_{\text{sol}})= 2.2>0$ implying that our solution is stable.
\paragraph*{Conclusion and remarks} The Schwinger effect in de Sitter spacetime is now a topical playground for cosmologists. However, very few was done regarding the role these pairs may play on the electromagnetic sector of the theory. In this letter, we completed the picture of the Schwinger effect by studying the backreaction of the Schwinger pairs to an electromagnetic field.  We first presented the basic setup for studying backreaction, then solved the Einstein-Maxwell equation numerically and investigated the parameter space. Our major conclusions are that the backreaction of fermion pairs created decrease or unaffect the electric field for all the parameters we considered. This corresponds to the screening of the fermions to the photon field. Conversely, for light bosons, we reported a solution where the electric field is enhanced. It corresponds to an anti-screening of the bosons to the photon field.
\par This regime where the amplification of the electromagnetic field occurs corresponds to a regime of infrared-hyperconductivity (IR-HC) \cite{Frob:2014zka,Kobayashi:2014zza}, where decreasing the electric field increases the resulting created pairs. In Ref.~\cite{Bavarsad:2016cxh}, the detailed regime of IR-HC in the D-dimension boson case is given and happened for $\left(\frac{eE}{H^2}\right)^2 \in (\frac{m^2}{H^2}, \frac{2D-3}{4}+\epsilon)$, with $0<\epsilon \ll 1$ for $D=2,3$ and $\epsilon$ positive but unbounded for $D=4$. Our solution with $\mu=0$ corresponds to cases where infrared-hyperconductivity is maximal.
\par We considered the 1 + 1 D case so that the physics becomes apparent but the generalization to higher dimension is the straightforward next step. Observe that for the 4D problem, equation (3.4) of \cite{Kobayashi:2014zza}, without the kinetic coupling, corresponds exactly to the 2D. Furthermore, the induced current has been found to have the same qualitative behavior up to corrections coming from dimension. Hence we expect our qualitative results: screening and anti-screening of the photon field to hold in higher dimensions.
\par Another direction worth to investigate would be to consider the full system $\{$Schwinger created particles, photon field$\}$, without the local constant field approximation used in this letter and deal with spatial variations of the electric field for the Schwinger pair creation. Aiming at doing that a technical point needs to be emphasized. While for the constant electric field, we replaced the Fourier transformed electric field \ref{eq:electricF} in the equation (96) of \cite{Stahl:2015gaa} to obtain \ref{eq:current}, in the case of space dependent electric field, one has to replace in the Schwinger current the electric field in position space \ref{eq:electric} and then perform a Fourier transform of the Schwinger current. We argue that to study such backreaction problems, the approach presented in \cite{Stahl:2015gaa} will not be sufficient and it will be required to study the Schwinger effect in de Sitter space with methods more suited for involved numerical investigation. For instance real-time lattice simulation techniques have been applied successfully in \cite{Kasper:2014uaa} to study the backreaction problem in flat spacetime. More techniques are also described in \cite{Gelis:2015kya} but their generalization to de Sitter spacetime are beyond the scope of this letter.
\par Magnetic fields are ubiquitous in the Universe but their origin is still a mystery \cite{Grasso:2000wj}. Their generation mechanism can be roughly divided into two categories: primoridial ones (a review and a recent  example: \cite{Subramanian:2015lua,Ferreira:2013sqa}) which happened before recombination and astrophysical ones (a review and a recent example: \cite{Brandenburg:2004jv,Durrive:2015cja}) happening after recombination. In this letter, we investigated an alternative possibility of amplifying the electromagnetic field with the help of the backreaction of the Schwinger particles created during inflation. For primordial scenario during inflation, to enhance an electromagnetic field, one needs to break the conformal invariance of Maxwell theory. Usually this is done by introducing a non-canonical kinetic term or by adding a mass to the photon field. Those magnetogenesis scenarii are known to suffer problems such as ghosts, the strong coupling problem and the backreaction problem. However the backreaction to the electric field was never investigated before and could change drastically the dynamics as we described already in 1 + 1 D (see also reference \cite{,Yokoyama:2015wws} for a brief review of inflationary magnetogenesis and its possible connection to Schwinger effect in curved spacetime). Furthermore if the Schwinger pairs are light enough, the electromagnetic field could then be enhanced without any other mechanism. In the future, we will study the same mechanism with the presence of magnetic field in the 4D case to show  whether it is relevant to the primordial magnetogenesis and applicable to the inflationary era.

\begin{acknowledgments}
CS is supported by the Erasmus Mundus Joint Doctorate Program by Grant Number 2013-1471 from the EACEA of the European
Commission.
\end{acknowledgments}


\begin{thebibliography}{25}
\expandafter\ifx\csname natexlab\endcsname\relax\def\natexlab#1{#1}\fi
\expandafter\ifx\csname bibnamefont\endcsname\relax
  \def\bibnamefont#1{#1}\fi
\expandafter\ifx\csname bibfnamefont\endcsname\relax
  \def\bibfnamefont#1{#1}\fi
\expandafter\ifx\csname citenamefont\endcsname\relax
  \def\citenamefont#1{#1}\fi
\expandafter\ifx\csname url\endcsname\relax
  \def\url#1{\texttt{#1}}\fi
\expandafter\ifx\csname urlprefix\endcsname\relax\def\urlprefix{URL }\fi
\providecommand{\bibinfo}[2]{#2}
\providecommand{\eprint}[2][]{\url{#2}}

\bibitem[{\citenamefont{Sauter}(1931)}]{Sauter1931}
\bibinfo{author}{\bibfnamefont{F.}~\bibnamefont{Sauter}}, \bibinfo{journal}{Z.
  Phys.} \textbf{\bibinfo{volume}{69}}, \bibinfo{pages}{742}
  (\bibinfo{year}{1931}).

\bibitem[{\citenamefont{Heisenberg and Euler}(1936)}]{Heisenberg1936}
\bibinfo{author}{\bibfnamefont{W.}~\bibnamefont{Heisenberg}} \bibnamefont{and}
  \bibinfo{author}{\bibfnamefont{H.}~\bibnamefont{Euler}}, \bibinfo{journal}{Z.
  Phys.} \textbf{\bibinfo{volume}{98}}, \bibinfo{pages}{714}
  (\bibinfo{year}{1936}).

\bibitem[{\citenamefont{Schwinger}(1954)}]{Schwinger1954A}
\bibinfo{author}{\bibfnamefont{J.}~\bibnamefont{Schwinger}},
  \bibinfo{journal}{Phys. Rev.} \textbf{\bibinfo{volume}{93}},
  \bibinfo{pages}{615} (\bibinfo{year}{1954}).

\bibitem[{\citenamefont{Ruffini et~al.}(2010)\citenamefont{Ruffini,
  Vereshchagin, and Xue}}]{Ruffini:2009hg}
\bibinfo{author}{\bibfnamefont{R.}~\bibnamefont{Ruffini}},
  \bibinfo{author}{\bibfnamefont{G.}~\bibnamefont{Vereshchagin}},
  \bibnamefont{and} \bibinfo{author}{\bibfnamefont{S.-S.} \bibnamefont{Xue}},
  \bibinfo{journal}{Phys. Rept.} \textbf{\bibinfo{volume}{487}},
  \bibinfo{pages}{1} (\bibinfo{year}{2010}), \eprint{0910.0974}.

\bibitem[{\citenamefont{Gelis and Tanji}(2016)}]{Gelis:2015kya}
\bibinfo{author}{\bibfnamefont{F.}~\bibnamefont{Gelis}} \bibnamefont{and}
  \bibinfo{author}{\bibfnamefont{N.}~\bibnamefont{Tanji}},
  \bibinfo{journal}{Prog. Part. Nucl. Phys.} \textbf{\bibinfo{volume}{87}},
  \bibinfo{pages}{1} (\bibinfo{year}{2016}), \eprint{1510.05451}.

\bibitem[{\citenamefont{Baumann and McAllister}(2015)}]{Baumann:2014nda}
\bibinfo{author}{\bibfnamefont{D.}~\bibnamefont{Baumann}} \bibnamefont{and}
  \bibinfo{author}{\bibfnamefont{L.}~\bibnamefont{McAllister}},
  \emph{\bibinfo{title}{{Inflation and String Theory}}}
  (\bibinfo{publisher}{Cambridge University Press}, \bibinfo{year}{2015}), ISBN
  \bibinfo{isbn}{9781107089693, 9781316237182}, \eprint{1404.2601},
  \urlprefix\url{http://inspirehep.net/record/1289899/files/arXiv:1404.2601.pdf}.

\bibitem[{\citenamefont{Hawking}(1975)}]{Hawking:1974sw}
\bibinfo{author}{\bibfnamefont{S.~W.} \bibnamefont{Hawking}},
  \bibinfo{journal}{Commun. Math. Phys.} \textbf{\bibinfo{volume}{43}},
  \bibinfo{pages}{199} (\bibinfo{year}{1975}), \bibinfo{note}{[,167(1975)]}.

\bibitem[{\citenamefont{Unruh}(1976)}]{Unruh:1976db}
\bibinfo{author}{\bibfnamefont{W.~G.} \bibnamefont{Unruh}},
  \bibinfo{journal}{Phys. Rev.} \textbf{\bibinfo{volume}{D14}},
  \bibinfo{pages}{870} (\bibinfo{year}{1976}).

\bibitem[{\citenamefont{Bucher}(2015)}]{Bucher:2015sbn}
\bibinfo{author}{\bibfnamefont{M.}~\bibnamefont{Bucher}}
  (\bibinfo{collaboration}{Planck}), \bibinfo{journal}{Nucl. Part. Phys. Proc.}
  \textbf{\bibinfo{volume}{267-269}}, \bibinfo{pages}{245}
  (\bibinfo{year}{2015}).

\bibitem[{\citenamefont{Fröb et~al.}(2014)\citenamefont{Fröb, Garriga, Kanno,
  Sasaki, Soda, Tanaka, and Vilenkin}}]{Frob:2014zka}
\bibinfo{author}{\bibfnamefont{M.~B.} \bibnamefont{Fröb}},
  \bibinfo{author}{\bibfnamefont{J.}~\bibnamefont{Garriga}},
  \bibinfo{author}{\bibfnamefont{S.}~\bibnamefont{Kanno}},
  \bibinfo{author}{\bibfnamefont{M.}~\bibnamefont{Sasaki}},
  \bibinfo{author}{\bibfnamefont{J.}~\bibnamefont{Soda}},
  \bibinfo{author}{\bibfnamefont{T.}~\bibnamefont{Tanaka}}, \bibnamefont{and}
  \bibinfo{author}{\bibfnamefont{A.}~\bibnamefont{Vilenkin}},
  \bibinfo{journal}{JCAP} \textbf{\bibinfo{volume}{1404}}, \bibinfo{pages}{009}
  (\bibinfo{year}{2014}), \eprint{1401.4137}.

\bibitem[{\citenamefont{Kobayashi and Afshordi}(2014)}]{Kobayashi:2014zza}
\bibinfo{author}{\bibfnamefont{T.}~\bibnamefont{Kobayashi}} \bibnamefont{and}
  \bibinfo{author}{\bibfnamefont{N.}~\bibnamefont{Afshordi}},
  \bibinfo{journal}{JHEP} \textbf{\bibinfo{volume}{10}}, \bibinfo{pages}{166}
  (\bibinfo{year}{2014}), \eprint{1408.4141}.

\bibitem[{\citenamefont{Fischler et~al.}(2015)\citenamefont{Fischler, Nguyen,
  Pedraza, and Tangarife}}]{Fischler:2014ama}
\bibinfo{author}{\bibfnamefont{W.}~\bibnamefont{Fischler}},
  \bibinfo{author}{\bibfnamefont{P.~H.} \bibnamefont{Nguyen}},
  \bibinfo{author}{\bibfnamefont{J.~F.} \bibnamefont{Pedraza}},
  \bibnamefont{and}
  \bibinfo{author}{\bibfnamefont{W.}~\bibnamefont{Tangarife}},
  \bibinfo{journal}{Phys. Rev.} \textbf{\bibinfo{volume}{D91}},
  \bibinfo{pages}{086015} (\bibinfo{year}{2015}), \eprint{1411.1787}.

\bibitem[{\citenamefont{Stahl et~al.}(2016)\citenamefont{Stahl, Strobel, and
  Xue}}]{Stahl:2015gaa}
\bibinfo{author}{\bibfnamefont{C.}~\bibnamefont{Stahl}},
  \bibinfo{author}{\bibfnamefont{E.}~\bibnamefont{Strobel}}, \bibnamefont{and}
  \bibinfo{author}{\bibfnamefont{S.-S.} \bibnamefont{Xue}},
  \bibinfo{journal}{Phys. Rev.} \textbf{\bibinfo{volume}{D93}},
  \bibinfo{pages}{025004} (\bibinfo{year}{2016}), \eprint{1507.01686}.

\bibitem[{\citenamefont{Bavarsad et~al.}(2016)\citenamefont{Bavarsad, Stahl,
  and Xue}}]{Bavarsad:2016cxh}
\bibinfo{author}{\bibfnamefont{E.}~\bibnamefont{Bavarsad}},
  \bibinfo{author}{\bibfnamefont{C.}~\bibnamefont{Stahl}}, \bibnamefont{and}
  \bibinfo{author}{\bibfnamefont{S.-S.} \bibnamefont{Xue}}
  (\bibinfo{year}{2016}), \eprint{1602.06556}.

\bibitem[{\citenamefont{Hayashinaka et~al.}(2016)\citenamefont{Hayashinaka,
  Fujita, and Yokoyama}}]{Hayashinaka:2016qqn}
\bibinfo{author}{\bibfnamefont{T.}~\bibnamefont{Hayashinaka}},
  \bibinfo{author}{\bibfnamefont{T.}~\bibnamefont{Fujita}}, \bibnamefont{and}
  \bibinfo{author}{\bibfnamefont{J.}~\bibnamefont{Yokoyama}}
  (\bibinfo{year}{2016}), \eprint{1603.04165}.

\bibitem[{\citenamefont{Hayashinaka and Yokoyama}(2016)}]{Hayashinaka:2016dnt}
\bibinfo{author}{\bibfnamefont{T.}~\bibnamefont{Hayashinaka}} \bibnamefont{and}
  \bibinfo{author}{\bibfnamefont{J.}~\bibnamefont{Yokoyama}}
  (\bibinfo{year}{2016}), \eprint{1603.06172}.

\bibitem[{\citenamefont{Kluger et~al.}(1991)\citenamefont{Kluger, Eisenberg,
  Svetitsky, Cooper, and Mottola}}]{Kluger:1991ib}
\bibinfo{author}{\bibfnamefont{Y.}~\bibnamefont{Kluger}},
  \bibinfo{author}{\bibfnamefont{J.~M.} \bibnamefont{Eisenberg}},
  \bibinfo{author}{\bibfnamefont{B.}~\bibnamefont{Svetitsky}},
  \bibinfo{author}{\bibfnamefont{F.}~\bibnamefont{Cooper}}, \bibnamefont{and}
  \bibinfo{author}{\bibfnamefont{E.}~\bibnamefont{Mottola}},
  \bibinfo{journal}{Phys. Rev. Lett.} \textbf{\bibinfo{volume}{67}},
  \bibinfo{pages}{2427} (\bibinfo{year}{1991}).

\bibitem[{\citenamefont{Kluger et~al.}(1992)\citenamefont{Kluger, Eisenberg,
  Svetitsky, Cooper, and Mottola}}]{PhysRevD.45.4659}
\bibinfo{author}{\bibfnamefont{Y.}~\bibnamefont{Kluger}},
  \bibinfo{author}{\bibfnamefont{J.~M.} \bibnamefont{Eisenberg}},
  \bibinfo{author}{\bibfnamefont{B.}~\bibnamefont{Svetitsky}},
  \bibinfo{author}{\bibfnamefont{F.}~\bibnamefont{Cooper}}, \bibnamefont{and}
  \bibinfo{author}{\bibfnamefont{E.}~\bibnamefont{Mottola}},
  \bibinfo{journal}{Phys. Rev. D} \textbf{\bibinfo{volume}{45}},
  \bibinfo{pages}{4659} (\bibinfo{year}{1992}),
  \urlprefix\url{http://link.aps.org/doi/10.1103/PhysRevD.45.4659}.

\bibitem[{\citenamefont{Ruffini et~al.}(2003)\citenamefont{Ruffini, Vitagliano,
  and Xue}}]{Ruffini:2003cr}
\bibinfo{author}{\bibfnamefont{R.}~\bibnamefont{Ruffini}},
  \bibinfo{author}{\bibfnamefont{L.}~\bibnamefont{Vitagliano}},
  \bibnamefont{and} \bibinfo{author}{\bibfnamefont{S.~S.} \bibnamefont{Xue}},
  \bibinfo{journal}{Phys. Lett.} \textbf{\bibinfo{volume}{B559}},
  \bibinfo{pages}{12} (\bibinfo{year}{2003}), \eprint{astro-ph/0302549}.

\bibitem[{\citenamefont{Ruffini et~al.}(2007)\citenamefont{Ruffini,
  Vereshchagin, and Xue}}]{Ruffini:2007jm}
\bibinfo{author}{\bibfnamefont{R.}~\bibnamefont{Ruffini}},
  \bibinfo{author}{\bibfnamefont{G.~V.} \bibnamefont{Vereshchagin}},
  \bibnamefont{and} \bibinfo{author}{\bibfnamefont{S.~S.} \bibnamefont{Xue}},
  \bibinfo{journal}{Phys. Lett.} \textbf{\bibinfo{volume}{A371}},
  \bibinfo{pages}{399} (\bibinfo{year}{2007}), \eprint{0706.4363}.
%\cite{Kasper:2014uaa}
\bibitem{Kasper:2014uaa}
  V.~Kasper, F.~Hebenstreit and J.~Berges,
  %``Fermion production from real-time lattice gauge theory in the classical-statistical regime,''
  Phys.\ Rev.\ D {\bf 90} no.2,  025016 (2014),
1403.4849.
  
  \bibitem[{\citenamefont{Grasso and Rubinstein}(2001)}]{Grasso:2000wj}
\bibinfo{author}{\bibfnamefont{D.}~\bibnamefont{Grasso}} \bibnamefont{and}
  \bibinfo{author}{\bibfnamefont{H.~R.} \bibnamefont{Rubinstein}},
  \bibinfo{journal}{Phys. Rept.} \textbf{\bibinfo{volume}{348}},
  \bibinfo{pages}{163} (\bibinfo{year}{2001}), \eprint{astro-ph/0009061}.

\bibitem[{\citenamefont{Subramanian}(2015)}]{Subramanian:2015lua}
\bibinfo{author}{\bibfnamefont{K.}~\bibnamefont{Subramanian}}
  (\bibinfo{year}{2015}), \eprint{1504.02311}.

\bibitem[{\citenamefont{Ferreira et~al.}(2013)\citenamefont{Ferreira, Jain, and
  Sloth}}]{Ferreira:2013sqa}
\bibinfo{author}{\bibfnamefont{R.~J.~Z.} \bibnamefont{Ferreira}},
  \bibinfo{author}{\bibfnamefont{R.~K.} \bibnamefont{Jain}}, \bibnamefont{and}
  \bibinfo{author}{\bibfnamefont{M.~S.} \bibnamefont{Sloth}},
  \bibinfo{journal}{JCAP} \textbf{\bibinfo{volume}{1310}}, \bibinfo{pages}{004}
  (\bibinfo{year}{2013}), \eprint{1305.7151}.

\bibitem[{\citenamefont{Brandenburg and
  Subramanian}(2005)}]{Brandenburg:2004jv}
\bibinfo{author}{\bibfnamefont{A.}~\bibnamefont{Brandenburg}} \bibnamefont{and}
  \bibinfo{author}{\bibfnamefont{K.}~\bibnamefont{Subramanian}},
  \bibinfo{journal}{Phys. Rept.} \textbf{\bibinfo{volume}{417}},
  \bibinfo{pages}{1} (\bibinfo{year}{2005}), \eprint{astro-ph/0405052}.

\bibitem[{\citenamefont{Durrive and Langer}(2015)}]{Durrive:2015cja}
\bibinfo{author}{\bibfnamefont{J.-B.} \bibnamefont{Durrive}} \bibnamefont{and}
  \bibinfo{author}{\bibfnamefont{M.}~\bibnamefont{Langer}},
  \bibinfo{journal}{Mon. Not. Roy. Astron. Soc.}
  \textbf{\bibinfo{volume}{453}}, \bibinfo{pages}{345} (\bibinfo{year}{2015}),
  \eprint{1506.08177}.

\bibitem{Yokoyama:2015wws}
  J.~Yokoyama,
  %``Issues on the inflationary magnetogenesis,''
  Comptes Rendus Physique {\bf 16} (2015) no.10,  1018.
  

\end{thebibliography}
\end{document}